\documentclass[epj]{svjour}
\newcommand*{\softfont}[1]{\textsf{#1}}

\newcommand*{\ddhep}{\softfont{DD4hep}}
\newcommand*{\edmhep}{\softfont{EDM4hep}}
\newcommand*{\keyhep}{\softfont{Key4hep}}

\newcommand*{\ilcsoft}{\softfont{iLCSoft}}
\newcommand*{\marlin}{\softfont{Marlin}}
\newcommand*{\gaudi}{\softfont{Gaudi}}
\newcommand*{\rootsoft}{\softfont{ROOT}}
\newcommand*{\geant}{\softfont{Geant4}}
\newcommand*{\pandora}{\softfont{PandoraPFA}}

\newcommand*{\delphes}{\softfont{Delphes}}
\newcommand*{\acts}{\softfont{ACTS}}
\newcommand*{\podio}{\softfont{PODIO}}
\newcommand*{\lcio}{\softfont{LCIO}}

\newcommand*{\keyCore}{\softfont{k4FWCore}}
\newcommand*{\keyDelphes}{\softfont{k4SimDelphes}}
\newcommand*{\keyGen}{\softfont{k4Gen}}
\newcommand*{\keyGeant}{\softfont{k4SimGeant4}}
\newcommand*{\keyRecCalo}{\softfont{k4RecCalorimeter}}

\newcommand*{\fccsw}{\softfont{FCCSW}}
\newcommand*{\fccedm}{\softfont{FCC-edm}}

\newcommand*{\fcc}{\softfont{FCC}}
\newcommand*{\fcchh}{\softfont{FCC-hh}}
\newcommand*{\fccee}{\softfont{FCC-ee}}
\newcommand*{\fccdetectors}{\softfont{FCCDetectors}}
\newcommand*{\lego}{\softfont{LEGO}}
\newcommand*{\yaml}{\softfont{YAML}}
\newcommand*{\spack}{\softfont{Spack}}
\newcommand*{\clue}{\softfont{CLUE}}

\newcommand*{\rnd}{research and development}

\usepackage{graphics}
\usepackage{listings}
\usepackage[pagewise]{lineno}
\begin{document}
\title{Key4hep, a framework for future HEP experiments and its use in FCC}
\author{Gerardo Ganis\inst{1} \and Cl\'ement Helsens\inst{1} \and Valentin V\"olkl\inst{1}
}                     
\offprints{}          
\institute{CERN}
\date{{\it (Submitted to EPJ+ special issue: A future Higgs and Electroweak factory (FCC): Challenges towards
discovery, Focus on FCC-ee)}}
%
\abstract{
The road map to the FCC Feasibility Study Report, for submission to the next Update of the European Strategy for Particle Physics, will require detailed simulation and advanced reconstruction algorithms to explore and maximise the physics reach of proposed detector solutions. The optimisation process will require maximal flexibility in changing detector geometries, materials and sensitive areas, and efficient tools to quantify the overall performance. To synergise such developments the CEPC, CLIC, FCC, ILC and SCT communities have engaged in the commissioning of a `Turnkey Software Stack' (\keyhep{}), which would provide all the necessary ingredients, from simulation to analysis, for future experiments. This approach is based on the positive experience of the linear collider projects ILC and CLIC, which have developed and used a common software stack (\ilcsoft{}) over the last decade. \keyhep{} aims to cover most, if not all, future linear and circular machines colliding leptons (electrons, muons), and hadrons. The common software ecosystem will facilitate writing specific components for experiments ensuring coherency and maximising the re-use of established solutions. 
Project-specific software frameworks will require adaptation to fully profit from the common software base. 
In this essay we present the status and plans for re-framing the FCC software framework, \fccsw{}, around \keyhep{} and discuss the challenges associated with the transition.
%
} 
\maketitle
\section{Introduction}\label{sec:intro}

The road map to deepen the understanding of the physics potential of the Future Circular Collider (FCC), due for the next Update of the European Strategy for Particle Physics (UESPP), will require detailed simulation and advanced reconstruction algorithms to explore and maximise the physics reach of proposed detector solutions. The results submitted to the 2019~UESPP were obtained using \fccsw{}, an experiment software based on the LHC experience integrated with the results of developing common software projects, such as \ddhep{}, which served the initial needs of the FCC community~\cite{fccswchep} well. The next phase of the studies will require further developments of the software, in particular concerning the flexibility in sub-detector choices, including the simulation and reconstruction of their response.
The interplay between reconstruction algorithms and detector geometry, for example in particle flow clustering, means that the detector hardware cannot be developed and designed independently from the software. At the same time, developing and validating sophisticated algorithms, including accounting for a large number of edge cases, requires a significant amount of resources.
Thus the communities for future experiments -- CEPC~\cite{cepc}, CLIC~\cite{Charles:2018vfv}, FCC~\cite{Abada:2019lih}, ILC~\cite{Bambade:2019fyw}, SCT~\cite{sct_summary}, STC~\cite{Luo:2019xqt}
-- came to the agreement that the development of a common software solution would benefit everyone~\cite{kickBO,kickHK}.
Software development is a collaborative effort, and one way \keyhep{} facilitates writing specific components for experiments is by maximising the re-use of established solutions and packages to benefit from existing community developments, for example, \rootsoft{}~\cite{ROOT}, \geant{}~\cite{Allison:2006ve}, \ddhep{}~\cite{Gaede:2020tui}, \gaudi{}~\cite{Clemencic:2017qgi} and \podio{}~\cite{Gaede:2020frr}. By using and contributing to other experiment-independent software packages, an `ecosystem' of HEP software is fostered.

In this essay we describe the proposal of the turnkey software stack, we outline the requirements and planned ingredients, and showcase the evolution of \fccsw{} towards this common software solution. We focus on the functionality provided by the ecosystem because we believe that matching the needs of the FCC project is the real challenge ahead. We expect other aspects, such as the one of computing performance, which will certainly become very important for the FCC operations, to be addressed, when relevant, at the level of the components that \keyhep{} brings together.

\section{\keyhep{}: the Turnkey Software Stack for Future Colliders}\label{sec:key4hep}

The turnkey software stack, nicknamed \keyhep{}, aims to encompass all the
libraries needed for generation, simulation, reconstruction, and analysis of events at the participating future colliders.
The project effectively started in February 2020 after two kick-off events which took place in Bologna~\cite{kickBO} and Hong-Kong~\cite{kickHK}. A more detailed description of the vision and goals of the project can be found in Ref.~\cite{keyhepchep}. In this section we will recall the main ingredients.

\subsection{Ingredients}\label{sec:ingred}

A software ecosystem for HEP data processing comprises several components which need to be chosen carefully. The most important are the following:
\begin{enumerate}
    \item A data processing framework, which provides the structure for all dependent components. 
    Despite the fact that the linear collider community has used \marlin{}~\cite{Gaede:2006pj} for many years very successfully, the choice has fallen on \gaudi{} because it is adopted by LHC experiments and it has a large user and developer community, offering - or planning to offer - support for access to heterogeneous resources, different architectures, and task-oriented
    concurrency.\footnote{A selected number of features provided by \marlin{} and  currently not available in \gaudi{}, e.g. capability to automatically generate steering file templates, will be contributed back to \gaudi{} as part of the \keyhep{} project.}
    \item A geometry description tool; for this purpose all of the communities concerned
          already use \ddhep{}, which offers a complete
          detector description for simulation, reconstruction and analysis.\footnote{\ddhep{} is not only of interest for future experiments community but also for LHC's: indeed it will be used in production, as of run 3, by CMS and LHCb. This will entail a consolidation and development process highly beneficial for the future experiments, in particular with respect to aspects more relevant for running experiments, e.g. alignment and calibration. In addition, support for common digitisation modules is planned as part of the \rnd{} program AIDAInnova~\cite{AIDAInnova} with the aim of facilitating the translation of the energy deposits produced by the full simulation into something which can be used to reconstruct physics objects.}
    \item A common event data model (see Section~\ref{sec:edm4hep}).
     \item A common build infrastructure. For the ease of use by librarians and developers, one needs to be able to build any and all pieces of the stack with minimum effort. Most packages have a similar build system based on CMake and follow common good practices which already reduces the maintenance burden, but the large number of packages requires additional tooling to automate the installation and ensure a consistently linked set of packages with correct dependency resolution. The investigations of the HEP Software Foundation packaging working group have identified the scientific package manager \spack{}~\cite{Spack} as a suitable solution, and it has since been successfully used for prototype builds and established as the main build tool for \keyhep{} software \cite{spackchep}. In addition, not only does \spack{} allow sharing of the build results but also the build recipes with a wider community.
\end{enumerate}

\subsection{\edmhep{}: the Event Data Model}\label{sec:edm4hep}
The event data model defines the structures needed to describe the required event information in the persistent store. While, in principle, this can be different from what is used in the transient store seen by the data processing algorithms, and also from algorithm to algorithm, adopting the same event model everywhere reduces the need for conversions, enhances generality and improves inter-operability.

Since the beginning of the FCC physics potential studies, the choice has been for an event data model managed by \podio{}~\cite{Gaede:2020frr}, a toolkit that generates the data model implementation from templates. This allows for a separation of the high level description in \yaml{} files using plain-old-data simple types, and the low level persistency layer that can be optimised according to the back-end. Once the required classes are defined in the \yaml{} file, the \podio{} tool creates the source code automatically. For \keyhep{} the same technology has been chosen, with an event data model, referred to as \edmhep{}, initially based on the \lcio{} and \fccedm{} classes, and improved from there as needed. 

 An underlying assumption or challenge is that the same event data model can be used for all types of HEP experiments, in particular for hadron and lepton colliders, a case which is particularly applicable to the integrated \fcc{} programme, which has a lepton collision phase followed by a hadron collision phase, similar to what happened with {\sf LEP} and {\sf LHC}; this seems to be the case with \edmhep{} and it will be further discussed in Section~\ref{sec:fccsw}. 

\subsection{Development strategy}\label{sec:strategy}

During the kick-off meetings~\cite{kickBO,kickHK} it became clear that to maximise the chances of an early adoption a {\it deliver-early-deliver-often} approach needed to be adopted. Furthermore, early identification of key components, to be made quickly available in usable form, was required in order to plan and make progress with the evaluation process.

The activities started from the definition of the common event data model \edmhep{}, which defines a sort of common language underlying the inter-operation of the various components. Once a usable version of \edmhep{}\footnote{Currently version 0.3.2, https://github.com/key4hep/EDM4hep.} was available, the core component of the framework, \keyCore{}\footnote{Currently version 1.0pre06, https://github.com/key4hep/k4FWCore.}, which provides the interfaces and the data service, needed to be made available; \keyCore{} is largely based on the corresponding components of \fccsw{}. The operative availability of \edmhep{} and \keyCore{} enabled the development and integration of data processing components, in the form of \gaudi{} algorithms and tasks. 

It was soon realised that having a full workflow producing `usable' events in \edmhep{} format would allow acceleration of the process. For this purpose the choice has been to integrate \delphes{}~\cite{delphes}, a popular tool for super-fast parameterised detector response. The component \keyDelphes{}, derived from a similar one existing in \fccsw{}, provides a \keyhep{} interface to \delphes{} both in terms of standalone applications, running on top of any of the input formats supported by \delphes{} and a \gaudi{} algorithm. 

In parallel, to facilitate the transition and interplay with \ilcsoft{}, a \gaudi{}-based wrapper around \marlin{} processes, which are the equivalent of the \gaudi{} algorithms, has been developed, enabling the complete \ilcsoft{} reconstruction via \gaudi{}\footnote{Currently version 0.3.2, https://github.com/key4hep/k4LCIOReader.} to be run.
The picture here is completed by the provision of an on-the-fly converter \lcio{}--\edmhep{}--\lcio{}\footnote{Currently version 0.3.1, https://github.com/key4hep/k4MarlinWrapper.} enabling the possibility of running an algorithm developed for \marlin{} and \lcio{}, for example the flavor tagging code LCFIplus, inside a full \gaudi{} application.  By design, \edmhep{} allows a  one-to-one conversion to \lcio\ and back without loss of information. In addition, the \podio{}-generated interfaces allow simple ad-hoc user data extensions of the data model should that be required.   

The basic components described above enable the development of specific new elements or the adoption of existing ones --  a process which is now ongoing for generation, full simulation and reconstruction capabilities.

\section{\fccsw{} and \keyhep{}}\label{sec:fccsw}
The FCC software framework, \fccsw{}, has a lot of commonalities with \keyhep{}, stemming from the fact that they are both based on \gaudi{}.
The main differences are the event data model and the structure governing the components, which in \keyhep{} is flat while \fccsw{} features a hierarchical structure with sub-components organised in categories.

For \fccsw{} moving to \keyhep{} conceptually means using common components available from \keyhep{} as much as possible, but FCC-specific components will be kept under FCC responsibility. 

From the technical point of view, the following chronologically-ordered steps were identified: 
\begin{enumerate}
    \item Replacement of \fccedm{} with \edmhep{};
    \item Identification and migration of non-FCC-specific \fccsw{} components to \keyhep{};
    \item Completion of the ecosystem, either in \keyhep{} or in the components remaining in \fccsw{}
    \begin{enumerate}
        \item Addition of missing sub-components, sub-detectors; optimisation/replacement of full/fast simulation techniques, reconstruction algorithms, etc.
    \end{enumerate}
\end{enumerate}

\subsection{Current status}\label{sec:fccswstatus}

\subsubsection{Migration to \edmhep{}}\label{sec:fccswedm}

The replacement of \fccedm{} with \edmhep{} has been completed without any major problem for all components, i.e. including those specific to \fcchh{}. This means that all the data structures needed to describe hadronic collisions could be handled with the structures provided by \edmhep{}.

\subsubsection{General and \fcc{}-specific \fccsw{} components}\label{sec:fccswcomp}
As mentioned above, \keyhep{} and \fccsw{} are conceptually very similar. The \fccsw{} components are therefore natural candidates to be generalised and become parts of upstream \keyhep{}. Table~\ref{tab:fccswcomp} summarises the current status of the main components.

\begin{table}[h]\centering
  \caption{Original \fccsw{} components and their fate in the new structures, as well as additional information on the current progress in transitioning them to use \keyhep{} interfaces ("status") and whether they will be ultimately hosted by \keyhep{} ("migration")  . See text for more details about {\sf dual--readout}.\label{tab:fccswcomp}}
  \begin{tabular}{|l||l|l||l|l|}
  \hline
    old \fccsw{} (version~$\leq$~0.16) & \keyhep{} & new \fccsw{} & status & migration  \\
  \hline
  \hline
    {\sf FWCore}              &  \keyCore{} & & done & yes \\
    {\sf Sim/SimDelphesInterface} &  \keyDelphes{} & & done & yes  \\
    {\sf Generation}          &  & \keyGen{}  & done & under evaluation \\
    {\sf Sim}           & &  \keyGeant{}   & done & under evaluation \\
    {\sf Reconstruction/Rec[...]Calorimeter} &  & \keyRecCalo{} & done & under evaluation \\

    {\sf Reconstruction/RecDriftChamber} &  & to be determined & &  \\
    {\sf Detector} &  & \fccdetectors{} & done & no, \fcc{} specific \\ \hline
    & to be determined & {\sf dual--readout} & & under evaluation \\  \hline
  \end{tabular}
\end{table}

As can be seen from the table, in preparing the migration to \keyhep{}, a general restructuring of the \fccsw{} repository has taken place, resulting in a flat structure similar to that of \keyhep{}, with sub-repositories promoted to separate repositories. Some of the components, {\sf FWCore} and {\sf SimDelphesInterface}, have been polished, renamed and migrated, and are already extensively used. For some others the preparation work has been completed and they are being analysed for migration. The detector descriptions have moved to a separate repository, \fccdetectors{}, which will not be migrated, as it is \fcc{}-specific.
The {\sf dual--readout} module includes \geant{} simulation and reconstruction of the {\sf IDEA} dual readout calorimeter; it has always been hosted in the \fcc{} project as a separate component, i.e. never part of \fccsw{}.
Finally, the {\sf RecDriftChamber} module implemented an approximate reconstruction of the {\sf IDEA} drift chamber and will not be migrated, since a detailed simulation and reconstruction for such a detector is being developed in \keyhep{} by the {\sf IDEA} working groups~\cite{ideatrack}.  

\subsubsection{Completion of the ecosystem}\label{sec:fccswcompletion}
 The completion of the ecosystem consists firstly of identifying the missing pieces in the findings of the other \fcc{} working groups, in particular the Physics Performance and Detector Concepts groups; secondly in creating the appropriate conditions for the provision of the required software module.
 
 Although this step will only take place once, the existing code is properly re-framed within \keyhep{} or \fccsw{} following the new paradigm. Some developments have already been enabled and can serve as an example of these activities.
 The first example is the ongoing process of integration of the simulation and
 reconstruction code for the {\sf IDEA} drift chamber and dual-readout calorimeter. For both of these important and innovative detectors the {\sf IDEA}
 collaboration developed standalone programs which are now being reviewed for inclusion in \keyhep{}. The process is well advanced and required some modification of \edmhep{}.
 The second example relates to track-related reconstruction algorithms, such as vertex finders, jet clustering and alike. The \delphes{} tool has recently been improved with a detailed simulation of the reconstructed parameters of electrically charged particles, with an output similar to that of full simulation and reconstruction. This allows algorithms to be developed which can serve as an advanced starting point when the tool is available.   

The completion of the ecosystem also includes a set of interfaces with external packages which are typically being developed as part of general purpose \rnd{} programs. Examples of these packages, for which the investigation work has already started, are the advanced tracking reconstruction system \acts{} \cite{Gessinger:2020nne}, the particle flow package \pandora{} \cite{Marshall:2015rfa} and the clustering algorithm \clue{} \cite{cluepaper}. One of the main challenges affecting the integration process is ensuring that the event data model contains all of the information required.

\subsection{The plug-and-play or \lego{} approach to detector composition and simulation}\label{sec:fccswlego}
The optimisation of detector design to support the full physics potential of  \fccee{} adds the challenge of finding the best combination of detection technologies. The fact that at least two -- but possibly four -- interaction regions are planned, opens the way to more solutions, potentially fine-tuned on specific physics aspects.

An infrastructure supporting the possibility to easily interchange sub-detector components is required for a broad investigation of all solutions. The \ddhep{} and \gaudi{} frameworks chosen for detector description and workflow execution support this through a plugin system which allows switching of both subdetectors and workflow components at runtime. The process has already been applied  when adapting the liquid Argon calorimeter solution developed for \fcchh{} to the \fccee{} case. Modular simulation workflows that allow the use of different fast and full simulation components in the same job, are being implemented.  The extension and subsequent validation of this approach, which for obvious reasons has been nicknamed \lego{}, is the goal and challenge of this exercise. 

\section{Conclusions and Outlook}\label{sec:conclusions}

The turnkey software stack, \keyhep{}, aims to create a complete data processing ecosystem for the benefit of HEP experiments, with immediate application to future collider designs and physics potential studies. The ecosystem will be built on established solutions such as \rootsoft{}, \geant{}, \ddhep{} and \gaudi{}, and features a new
event data model \edmhep{} based on \podio{}.
\keyhep{} already provides enough components and modules to enable the migration of existing frameworks based on \gaudi{}. In this essay we described the status of the migration of \fccsw{} which, at the time of writing, is close to completion.
The \keyhep{} projects seem to be on the right path to find the equilibrium to serve the needs of the several participating projects well, with the appropriate balance between common infrastructure and project-specific developments, and a clear path to promote the latter to the common repositories.

The challenge ahead is to maintain the current momentum and to create the conditions to attract and support people who will develop and implement their tools in the \keyhep{} context, thereby creating a critical mass of the best tools to cover most of the needs of HEP experiments. This task includes the provision of sufficiently extensive documentation, training and adequate advertising. Ultimately this will depend on the capability to guarantee an adequate  workforce and funding.

\section*{Acknowledgements}  \label{sec:acknowledgements}
This work beneﬁted from support by the CERN Strategic R\&D Programme on Technologies for Future Experiments (https://cds.cern.ch/record/2649646/, CERN-OPEN-2018-006).

%

\begin{thebibliography}{}
%
%


\bibitem{fccswchep}
See for example J.~Cervantes \textit{et al.}, ``A software framework for FCC studies: status and plans,'' CHEP 2019, EPJ Web of Conferences 245, 05018 (2020), doi:10.1051/epjconf/202024505018

\bibitem{cepc}
CEPC Study Group, ``CEPC conceptual design report: Volume 1-accelerator,'' [arXiv:1809.00285 [hep-ex]].

\bibitem{Charles:2018vfv}
P.~N.~Burrows \textit{et al.}
``The Compact Linear Collider (CLIC) - 2018 Summary Report,''
doi:10.23731/CYRM-2018-002
[arXiv:1812.06018 [physics.acc-ph]].

\bibitem{Abada:2019lih}
A.~Abada \textit{et al.}
``FCC Physics Opportunities: Future Circular Collider Conceptual Design Report Volume 1,''
Eur. Phys. J. C79 (2019) no.~6, 474
doi:10.1140/epjc/s10052-019-6904-3

\bibitem{Bambade:2019fyw}
P.~Bambade \textit{et al.},
``The International Linear Collider: A Global Project,''
[arXiv:1903.01629 [hep-ex]].



\bibitem{sct_summary}
Super Charm--Tau Factory
https://ctd.inp.nsk.su/c-tau/

\bibitem{Luo:2019xqt}
Q.~Luo \textit{et al.},
``Progress of Conceptual Study for the Accelerators of a 2-7GeV Super Tau Charm Facility at China,''
proceedings of 10$^{th}$ International Particle Accelerator Conference, Melbourne, Australia, 19-24 May 2019, JACoW Publishing, doi:10.18429/JACoW-IPAC2019-MOPRB031

\bibitem{kickBO}
``Future Collider Software Workshop,'' 12-13 June 2019, Bologna, https://agenda.infn.it/event/19047

\bibitem{kickHK}
``Mini-workshop: Experiment/Detector - Software and Physics Requirements for e+e- Colliders,'' 16-17 January 2020, Hong Kong,  http://iasprogram.ust.hk/hep/2020/


\bibitem{ROOT}
I. Antcheva \textit{et al.}, ``ROOT: A C++ framework for petabyte data storage, statistical analysis and visualization,'' Comput.~Phys.~Commun. 180 (2009) 2499-2512, doi:10.1016/j.cpc.2009.08.005 .

\bibitem{Allison:2006ve}
J.~Allison \textit{et al.}, ``Geant4 developments and applications,''
IEEE Trans. Nucl. Sci. 53 (2006), 270
doi:10.1109/TNS.2006.869826

\bibitem{Gaede:2020tui}
F.~Gaede, M.~Frank, M.~Petric and A.~Sailer,
``DD4hep a community driven detector description for HEP,''
EPJ Web Conf. 245 (2020), 02004
doi:10.1051/epjconf/202024502004

\bibitem{Clemencic:2017qgi}
M.~Clemencic \textit{et al.},
``Gaudi evolution for future challenges,''
J. Phys. Conf. Ser. 898 (2017) no.4, 042044
doi:10.1088/1742-6596/898/4/042044

\bibitem{Gaede:2020frr}
F.~Gaede \textit{et al.}, 
``PODIO: recent developments in the Plain Old Data EDM toolkit,''
EPJ Web Conf. 245 (2020), 05024
doi:10.1051/epjconf/202024505024

\bibitem{keyhepchep}
A.~Sailer \textit{et al.}, ``Towards a Turnkey Software Stack for HEP Experiments,'' CHEP 2019, EPJ Web Conf. 245 (2020) 10002,
doi: 10.1051/epjconf/202024510002

\bibitem{Gaede:2006pj}
F.~Gaede,
``Marlin and LCCD: Software tools for the ILC,''
Nucl. Instrum. Meth. A 559 (2006), 177-180, 
doi:10.1016/j.nima.2005.11.138

\bibitem{AIDAInnova}
AIDAInnova, European Union’s Horizon 2020 (Integrating Activity), WP12, task 12.2, https://aidainnova.web.cern.ch/wp12 

\bibitem{Spack}
T.~Gamblin \textit{et al.}, ``The Spack Package Manager: Bringing Order to HPC Software Chaos,'' proceedings of Supercomputing 2015 (SC’15), Austin, Texas, November 15-20 2015, doi 10.1145/2807591.2807623 . 

\bibitem{spackchep}
V.~Volkl \textit{et al.}, ``Building HEP Software with Spack: Experiences from Pilot Builds for Key4hep and Outlook for LCG Releases,'' proceedings of vCHEP 2021, EPJ Web of Conferences 251, 03056 (2021), doi https://doi.org/10.1051/epjconf/202125103056 

\bibitem{delphes}
J.~de Favereau \textit{et al.}
``DELPHES 3, A modular framework for fast simulation of a generic collider experiment,'' JHEP 02 (2014), 057, doi:10.1007/JHEP02(2014)057
[arXiv:1307.6346 [hep-ex]].



\bibitem{ideatrack}
See for example, N.~De~Filippis, ``Status of the IDEA Drift Chamber Software,'' presented at FCC Software Meeting of 11 December 2020, 
https://indico.cern.ch/event/979157/

\bibitem{Gessinger:2020nne}
P.~Gessinger \textit{et al.},
``The Acts project: track reconstruction software for HL-LHC and beyond,''
EPJ Web Conf. 245 (2020), 10003
doi:10.1051/epjconf/202024510003

\bibitem{Marshall:2015rfa}
J.~S.~Marshall and M.~A.~Thomson,
``The Pandora Software Development Kit for Pattern Recognition,''
Eur. Phys. J. C 75 (2015) no.9, 439
doi:10.1140/epjc/s10052-015-3659-3
[arXiv:1506.05348 [physics.data-an]].

\bibitem{cluepaper}
M. Rovere \textit{et al.}, ``CLUE: A Fast Parallel Clustering Algorithm for High Granularity Calorimeters in High Energy Physics,'' Jan 2020,  [arXiv:2001.09761 [hep-ex]].










\end{thebibliography}
%

\end{document}